\documentclass[twocolumn,PRB,aps,psfig,preprintnumbers,superscriptaddress]{revtex4}
\usepackage{graphicx}
\usepackage{epstopdf}
\usepackage{float}
\usepackage{color}
\usepackage{amsmath}

\begin{document}

\title{Incommensurate and commensurate antiferromagnetic states  in CaMn$_2$As$_2$ and SrMn$_2$As$_2$ revealed by $^{75}$As NMR} 
\author{Q.-P. Ding}
\affiliation{Ames Laboratory, Iowa State University, Ames, Iowa 50011, USA}
\affiliation{Department of Physics and Astronomy, Iowa State University, Ames, Iowa 50011, USA}
\author{N. S. Sangeetha$\footnote[1]{Present address: Institute for Experimental Physics~IV, Ruhr University Bochum, 44801 Bochum, Germany}$}
\affiliation{Ames Laboratory, Iowa State University, Ames, Iowa 50011, USA}
\author{Abhishek Pandey$\footnote[2]{Present address: School of Physics, University of the Witwatersrand, Johannesburg, Gauteng 2050, South Africa}$}
\affiliation{Ames Laboratory, Iowa State University, Ames, Iowa 50011, USA}
\author{D. C. Johnston}
\affiliation{Ames Laboratory, Iowa State University, Ames, Iowa 50011, USA}
\affiliation{Department of Physics and Astronomy, Iowa State University, Ames, Iowa 50011, USA}
\author{Y. Furukawa}
\affiliation{Ames Laboratory, Iowa State University, Ames, Iowa 50011, USA}
\affiliation{Department of Physics and Astronomy, Iowa State University, Ames, Iowa 50011, USA}
\date{\today}

\begin{abstract}

   We carried out  $^{75}$As nuclear magnetic resonance (NMR) measurements on the trigonal  CaMn$_2$As$_2$ and SrMn$_2$As$_2$ insulators  exhibiting antiferromagnetic (AFM) ordered states below N\'eel temperatures $T_{\rm N}$ = 62 K and 120 K, respectively.
   In the paramagnetic state above $T_{\rm N}$, typical  quadrupolar-split $^{75}$As-NMR spectra were observed for both systems. 
   The $^{75}$As quadrupolar frequency  $\nu_{\rm Q}$ for CaMn$_2$As$_2$  decreases with decreasing temperature, while $\nu_{\rm Q}$ for SrMn$_2$As$_2$ increases, showing an opposite temperature dependence.  
    In the AFM state, the relatively sharp and distinct $^{75}$As NMR lines were observed in SrMn$_2$As$_2$  and the NMR spectra were shifted to lower fields for both magnetic fields $H$ $\parallel$ $c$ axis  and $H$ $\parallel$ $ab$ plane, suggesting that the internal fields $B_{\rm int}$ at the As site produced by the Mn ordered moments are nearly perpendicular to the external magnetic field direction. 
  No obvious distribution of $B_{\rm int}$ was observed in SrMn$_2$As$_2$, which clearly indicates a commensurate AFM state.
   In sharp contrast to SrMn$_2$As$_2$, broad and complex NMR spectra were observed in  CaMn$_2$As$_2$ in the AFM state, which clearly shows a distribution of $B_{\rm int}$ at the As site, indicating an incommensurate state. 
   From the analysis of the characteristic shape of the observed spectra, the AFM state of  CaMn$_2$As$_2$ was determined to be a two-dimensional incommensurate  state where Mn ordered moments are aligned in the $ab$ plane. 
  A possible origin for the different AFM states in the systems was discussed. 
     Both CaMn$_2$As$_2$ and SrMn$_2$As$_2$ show very large anisotropy in the nuclear spin-lattice relaxation rate 1/$T_1$  in the paramagnetic state.
 1/$T_1$ for $H$  $\parallel$ $ab$ is much larger than that for $H$ $\parallel$ $c$, indicating strong anisotropic AFM spin fluctuations in both compounds.

\end{abstract}

\maketitle

\begin{figure}[tb]
\includegraphics[width=8.5cm]{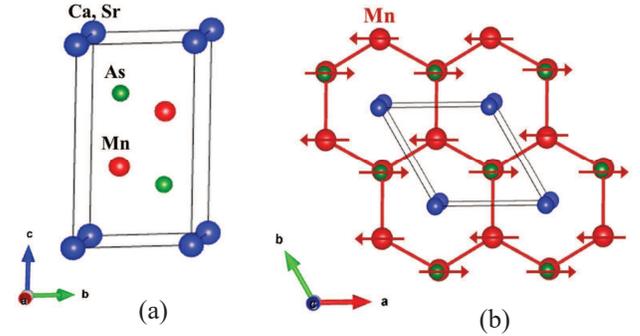} 
\caption{ Trigonal CaAl$_2$Si$_2$-type crystal structure of CaMn$_2$As$_2$ and SrMn$_2$As$_2$ in the hexagonal setting. 
(a) Outline of a unit cell containing one formula unit.
(b) Projection of the Mn sublattice onto the $ab$ plane with a slight $c$ axis component illustrating the corrugated Mn honeycomb lattice.
Red arrows represent the magnetic moments at the Mn sites for SrMn$_2$As$_2$.
  Black lines show a unit cell. 
 }
\label{fig:structure}
\end{figure}

 \section{Introduction} 

\    After the discovery of superconductivity (SC) in iron-based pnictides in 2008 \cite{Kamihara2008}, considerable experimental and theoretical attention has concentrated on  transition-metal pnictides  \cite{Johnston2010,Canfield2010,Stewart2011}. 
  Among them, the Mn compounds have been found to show a rich variety of magnetic properties with different crystal structures.
  BaMn$_2$As$_2$ with the body-centered tetragonal ThCr$_2$Si$_2$-type structure (space group $I4/mmm$), similar to the parent compound BaFe$_2$As$_2$, is a G-type collinear antiferromagnet with a N\'eel temperature of  625 K where the ordered Mn moments aligned along the tetragonal $c$ axis \cite{Singh2009}. 
  On the other hand,  (Ca,Sr)Mn$_2$P$_2$ \cite{Mewis1978,Sangeetha2021}, (Ca,Sr)Mn$_2$As$_2$ \cite{Mewis1978,Brechtel1978,Wang2011,Sangeetha2016}, (Ca,Sr)Mn$_2$Sb$_2$ \cite{Ratcliff II2009,Bridges2009,Sangeetha2018}, and CaMn$_2$Bi$_2$ \cite{Gibson2015} crystallize in the trigonal CaAl$_2$Si$_2$-type structure (space group $P\bar3m1$) shown in Fig.\ \ref{fig:structure}(a) where  the Mn ions form a triangular lattice bilayer which can be considered as a corrugated Mn honeycomb sublattice [see Fig.\ \ref{fig:structure}(b)] \cite{Gibson2015}.
    The honeycomb lattice structure is interesting as a spin system with spin frustration originating from  competing interactions \cite{Rastelli1979,Mazin2013,McNally2015}.
     Interestingly, recent theoretical studies and density-functional theory calculations have suggested that the CaAl$_2$Si$_2$-type  transition-metal pnictides might comprise a new family of magnetically-frustrated materials to study the potential superconducting mechanism \cite{Fouet2001,Zeng2017}.

    Magnetic susceptibility ($\chi$), electrical resistivity, and heat capacity measurements revealed that CaMn$_2$As$_2$ and SrMn$_2$As$_2$ are antiferromagnetic (AFM) insulators with $T_{\rm N}$ = 62(3) K and $T_{\rm N}$ = 120(2) K , respectively \cite {Sangeetha2016}. 
   In the AFM state, $\chi$ is nearly independent of temperature $T$ for magnetic fields $H$ $\parallel$ $c$ axis, while  $\chi$ for $H \parallel$~$ab$ plane decreases with decreasing temperature, indicating that the hexagonal $c$ axis is the hard axis and hence that the ordered Mn moments are aligned in the $ab$ plane. 
    However, $\chi_{ab}$ does not go to zero at low $T$ but attains a value $\chi_{ab}$($T\to$ 0) $\simeq$ 0.6 $\chi_c$($T_{\rm N}$), in contrast to the case of a simple collinear antiferromagnet with $H$ prallel to the ordering axis where  $\chi_{ab}$($T\to$ 0)$\to$ 0.
   If $H$ is perpendicular to the ordering axis, then $\chi_{ab}$($T\to$ 0)$\sim$  $\chi_{ab}$($T_{\rm N}$) \cite{Johnston2012,Johnston2015}.
    This suggests that the AFM structure could be either a collinear AFM with multiple domains aligned within the $ab$ plane or an intrinsic noncollinear structure with moments again aligned in the $ab$ plane \cite {Johnston2012,Johnston2015,Anand2015,Ryan2015}.     
    Subsequent single-crystal neutron-diffraction measurements on SrMn$_2$As$_2$ clearly demonstrated that the Mn moments (3.6 $\mu_{\rm B}$ at $T$  = 5 K) are ordered in a collinear N\'eel AFM phase with 180$^{\circ}$ AFM alignment between a moment and all nearest neighbor moments in the basal plane and also perpendicular to it [see Fig.~\ref{fig:structure}(b)] \cite{Das2016}.
   Therefore, in order to explain the finite $\chi_{ab}$($T \rightarrow$ 0), an occurrence of multiple AFM domains is suggested in SrMn$_2$As$_2$ \cite{Das2016, Sangeetha2016}.
   For CaMn$_2$As$_2$, the AFM magnetic structure has not been determined yet.

     In this paper we carried out  $^{75}$As nuclear magnetic resonance (NMR) measurements on trigonal CaMn$_2$As$_2$ and SrMn$_2$As$_2$ single crystals to investigate the static and dynamic magnetic properties from a microscopic point of view.  
   Our NMR measurements clearly detected the AFM phase transitions at  62 K and 120 K for CaMn$_2$As$_2$ and SrMn$_2$As$_2$, respectively.
    In the paramagnetic states, similar quadrupolar-split $^{75}$As NMR spectra were observed in both compounds. 
    However, quite different NMR spectra were detected in the AFM states.
   In CaMn$_2$As$_2$,  broad and complex NMR spectra were observed in the AFM state, which clearly indicates a distribution of internal fields $B_{\rm int}$ at the As site.
    From the analysis of the spectra, the AFM state of CaMn$_2$As$_2$ was determined to be a two-dimensional incommensurate  state where the Mn ordered moments are aligned in the $ab$ plane.    
   On the other hand, relatively sharp and  distinct NMR lines were observed in the AFM state in SrMn$_2$As$_2$. 
  The spectra indicate no obvious distribution of $B_{\rm int}$ in  SrMn$_2$As$_2$.
  These results indicate that, contrary to CaMn$_2$As$_2$, the AFM state in SrMn$_2$As$_2$ is commensurate.

\section{Experimental}

   The hexagonal-shaped single crystals of CaMn$_2$As$_2$ and SrMn$_2$As$_2$  for the NMR measurements were grown using Sn flux as reported in detail elsewhere \cite {Sangeetha2016}. 
   NMR measurements were carried out on $^{75}$As  (\textit{I}~=~3/2, $\gamma/2\pi$~=~7.2919 MHz/T, $Q$ =  0.29 barns)  using a homemade, phase-coherent, spin-echo pulse spectrometer. 
   The $^{75}$As-NMR spectra were obtained by sweeping the magnetic field $H$ at fixed frequencies $f$ = 51.1, 9.1 and  53.1 MHz.  
   The magnetic field was applied parallel to either the crystal $c$ axis ($H$ $||$ $c$) or to the $ab$ plane ($H$~$||$~$ab$). 
    In this paper, when we use $H$~$||$~$ab$, the magnetic field was applied along the edge direction of the hexagonal-shape crystal corresponding to the $a$ or $b$ axes,  except when we specify the direction in the $ab$ plane for angle-dependent NMR spectrum measurements in the $ab$ plane.  
The Tesla (T) is defined as 1 T =  10$^4$ Oe.  
   The $^{75}$As spin-lattice relaxation rate 1/$T_{\rm 1}$ was measured with a saturation-recovery method.
   $1/T_1$ at each $T$ was determined by fitting the nuclear magnetization $M$ versus time $t$  using the exponential functions $1-M(t)/M(\infty) = 0.1 e^ {-t/T_{1}} +0.9e^ {-6t/T_{1}}$ for $^{75}$As NMR,  where $M(t)$ and $M(\infty)$ are the nuclear magnetization at time $t$ after the saturation and the equilibrium nuclear magnetization at $t$ $\rightarrow$ $\infty$, respectively.    

\section{Results and discussion}
\subsection{CaMn$_2$As$_2$}
\subsubsection{$^{75}$As NMR in the paramagnetic state}

\begin{figure}[tb]
\includegraphics[width=8.5cm]{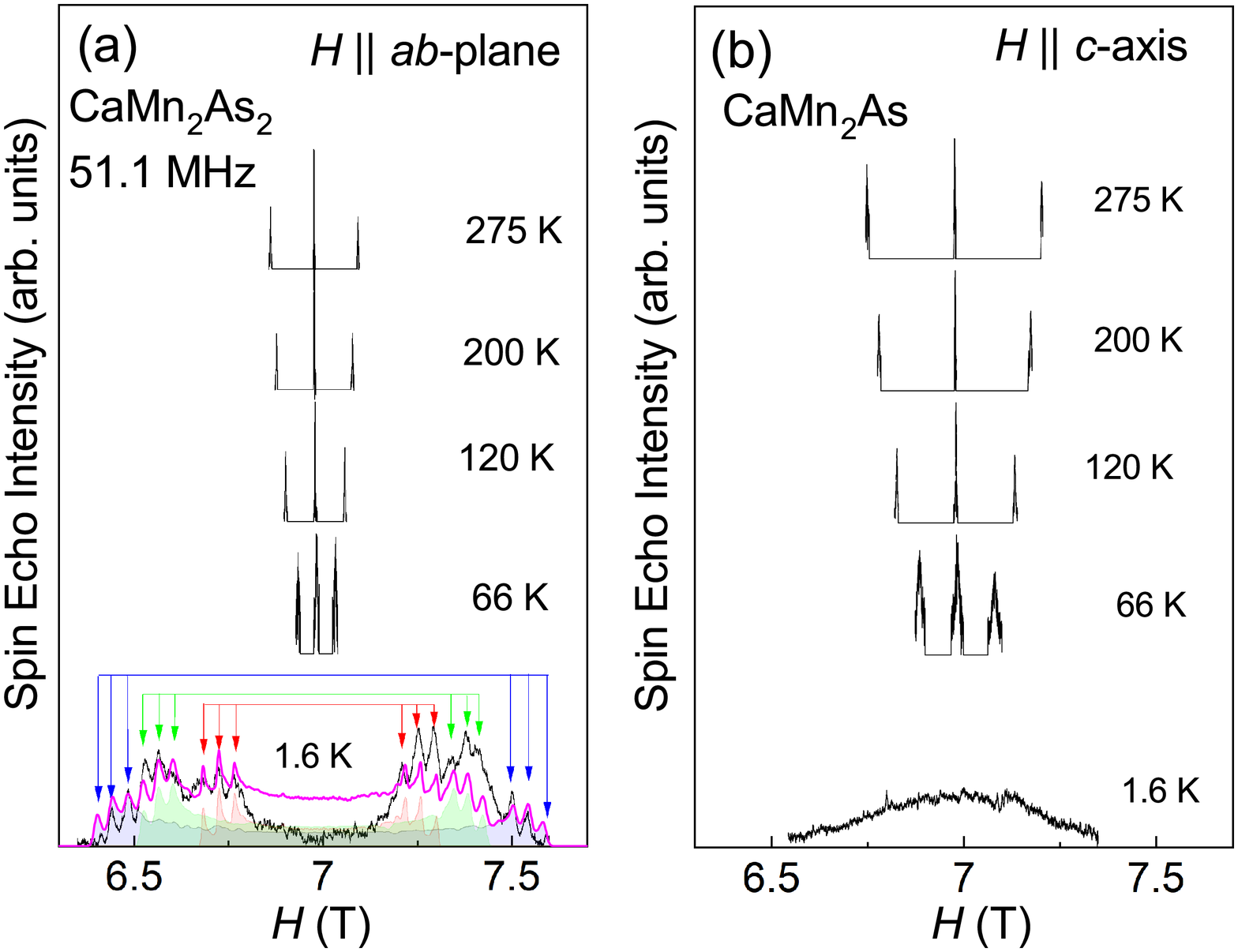} 
\caption{ Temperature variation of field-swept $^{75}$As-NMR spectra for a CaMn$_2$As$_2$ crystal at frequency $f$ = 51.1 MHz for magnetic field directions (a) $H$ $\parallel$ $ab$ plane and (b) $H$ $\parallel$ $c$ axis in the paramagnetic and antiferromagnetic states. 
 The spectra shown at  the bottoms in (a) and (b) were measured at 1.6 K. 
  The arrows in three different colors in (a) show the splitting of the $^{75}$As NMR spectra due to internal fields.  
   The blue, green and red areas are calculated NMR spectra based on the incommensurate AFM state for As1, As2 and As3 sites, respectively (see text). 
   The magenta curve is the sum of the three calculated spectra. 
}
\label{fig:CaSpec}
\end{figure}


    Figures\ \ref{fig:CaSpec}(a) and \ \ref{fig:CaSpec}(b) show the field-swept $^{75}$As-NMR spectra of CaMn$_2$As$_2$ for two magnetic field directions, $H$ $\parallel$ $ab$ plane and $H$ $\parallel$ $c$ axis, respectively, measured at various temperatures. 
       The typical spectrum for a nucleus with spin $I=3/2$ with Zeeman and quadrupolar interactions can be described by the nuclear spin Hamiltonian \cite{Slichter_book}, ${\cal H} = -\gamma_{\rm n}\hbar  (1+K) {\bf I} \cdot {\bf H} + \frac{h\nu_{\rm Q}}{6}[3I_Z^2-I^2 + \frac{1}{2}\eta(I_+^2 +I_-^2)]$,
where $\bf H$ is external field, $h$ is Planck's constant, and $K$ represents the NMR shift. 
The nuclear  quadrupole frequency for $I=3/2$ nuclei is given by $\nu_{\rm Q} = e^2QV_{\rm ZZ}/2h$, where $Q$ is the nuclear quadrupole moment, $V_{\rm ZZ}$ is the electric field gradient (EFG) at the As site, and $\eta$ is the asymmetry parameter of the EFG defined by $\frac{V_{XX} -V_{YY}}{V_{ZZ}}$ with $|V_{ZZ}|$$\geq$$|V_{YY}|$$\geq$$|V_{XX}|$.
   In the case of $I$ = 3/2,  when the Zeeman interaction is much greater than the quadrupolar interaction,  the NMR spectrum is composed of a central transition line  ($I_z$~=~1/2~$\leftrightarrow$~--1/2)  and  a pair of satellite lines shifted from the central transition line by $\pm\frac{1}{2} \nu_{\rm Q}(3\cos^2\theta -1 +\eta\sin^2\theta\cos2\phi)$ (for the transitions of $I_z$ = 3/2 $\leftrightarrow$ 1/2 and --3/2 $\leftrightarrow$ --1/2) \cite{Abragambook}.
 Here $\theta$ and $\phi$ are the polar and azimuthal angles between  the principal $Z$ axis of the EFG and the direction of $\bf H$, respectively.
   The observed $^{75}$As NMR spectra of CaMn$_2$As$_2$ in the paramagnetic state can be well reproduced by calculations from the above simple Hamiltonian with $\eta$ = 0 where the spectra for $H$ $||$ $c$ and $H$ $||$ $ab$ correspond to the case of $\theta$ = 0$^{\circ}$ and 90$^{\circ}$, respectively. 
   This indicates that the $Z$ axis of the EFG at the As site is parallel to the $c$ axis, consistent with the three-fold rotational symmetry around the $c$ axis at the As site.

\begin{figure}[tb]
\includegraphics[width=7.5cm]{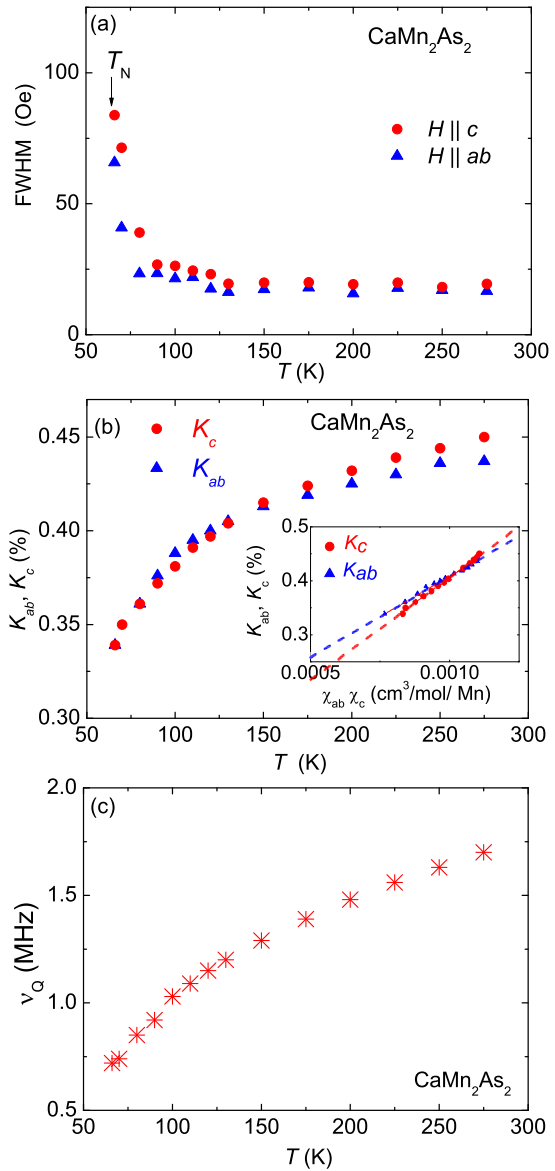} 
\caption{(a) Temperature dependence of the full width at half maximum (FWHM) of the central transition lines for $H$ $||$ $c$ and $H$ $||$ $ab$.
(b) Temperature dependence of the $^{75}$As NMR shift for $H$ $\parallel$ $c$ axis ($K_c$) and $H$ $\parallel$ $ab$ plane ($K_{ab}$).
Inset: $K$ vs magnetic susceptibility $\chi$ for the corresponding $ab$ and $c$ components of $K$ in CaMn$_2$As$_2$ with $T$ as an implicit parameter.  The dashed lines are linear fits.
(c) Temperature dependence of the quadrupole frequency $\nu_{\rm{Q}}$ for CaMn$_2$As$_2$.
}
\label{fig:CaSpec2}
\end{figure}

   The relatively sharp lines with the full width at half maximum (FWHM) of $\sim$ 20 and 17 Oe for $H$ $||$ $c$ and $H$ $||$~$ab$, respectively,  for the central transition lines are nearly independent of temperature above 150 K, as shown in Fig.\ \ref{fig:CaSpec2}(a).  
   Below 150 K, with decreasing temperature, the FWHM values start to increase gradually and rapidly increase below $\sim$ 70 K due to the AFM ordering at $T_{\rm N}$ = 62 K, indicating the second-order nature of the AFM phase transition in CaMn$_2$As$_2$.

    Figure~\ref{fig:CaSpec2}(b) shows the temperature dependence of the NMR shifts for $H$ $\parallel$ $c$ axis ($K_c$)  and $H \parallel ab$ plane ($K_{ab}$) for CaMn$_2$As$_2$.
   For both directions, $K$ decreases slightly with decreasing temperature from 300 K to $T_{\rm N}$, consistent with the temperature dependence of the magnetic susceptibility \cite{Sangeetha2016}, where a broad peak in  $\chi$ was observed around $T$ = 400 K. 
    The temperature dependences of $K$ and $\chi$ indicate short-range AFM order in the paramagnetic state \cite{Sangeetha2016}.   
    The NMR shift has contributions from the $T$-dependent spin part $K_{\rm spin}$ and a $T$-independent orbital part $K_0$. 
     $K_{\rm spin}$ is proportional to the spin susceptibility $\chi_{\rm spin}$ through the hyperfine coupling constant $A$ giving $K(T)=K_0+\frac{A}{N_{\rm A}}\chi_{\rm spin}(T)$, where $N_{\rm A}$ is Avogadro's number.
     The inset of Fig. \ref{fig:CaSpec2}(b)  plots $K_{ab}$ and $K_{c}$ for CaMn$_2$As$_2$ against the corresponding  $\chi_{ab}$  and  $\chi_c$, respectively, with $T$ as an implicit parameter. 
    All $K_{ab}$ and $K_{c}$ are seen to vary linearly with the corresponding $\chi$, and the hyperfine coupling constants are estimated to be $A_{c}$ = (--2.10 $\pm$ 0.03) T/$\mu_{\rm B}$, and $A_{ab}$ = (--1.66 $\pm$ 0.04) T/$\mu_{\rm B}$ for $H$ $\parallel$ $c$  and $H$ $\parallel$ $ab$, respectively.

\begin{figure}[tb]
\includegraphics[width=8.5cm]{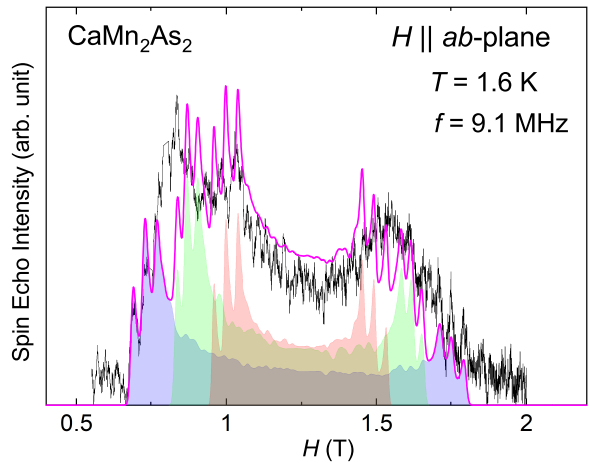} 
\caption{Field-swept $^{75}$As-NMR spectrum  measured at $T$~=~1.6~K and $f$ = 9.1 MHz for $H$ $\parallel$ $ab$ in the AFM state. 
The blue, green and red areas are calculated NMR spectra based on the incommensurate AFM state for As1, As2 and As3 sites, respectively. 
Here the parameters of $B_{\rm int}$ and $\nu_{\rm Q}$ used are the same as those estimated from the spectrum measured at 51.1 MHz. 
The magenta curve is the sum of the three calculated spectra.  
}
\label{fig:CaSpec3}
\end{figure}

   From the spectra shown in Figs.\ \ref{fig:CaSpec}(a) and \ref{fig:CaSpec}(b), we also extracted the temperature dependence of  $\nu_{\rm{Q}}$ for CaMn$_2$As$_2$, which is shown in Fig.\ \ref{fig:CaSpec2}(c). 
   $\nu_{\rm{Q}}$ decreases from 1.70 MHz at $T$ = 275 K to 0.70~MHz at 66 K.
    This is in contrast to the case of SrMn$_2$As$_2$ as will be shown below where $\nu_{\rm{Q}}$ increases with decreasing temperature.  


 \subsubsection{$^{75}$As NMR spectra in the antiferromagnetic state}

     Below $T_{\rm N}$ = 62 K, the NMR lines broaden suddenly and the spectrum could not be measured in the AFM state due to poor signal intensity, except at  the lowest temperature $T$ = 1.6 K of our $^4$He cryostat  where we were able to measure the spectra although the signal intensities were still weak.
     As shown at the bottom in Fig.~\ref{fig:CaSpec}(b), a single broad spectrum without clear splitting is observed for $H$ $||$ $c$. 
     On the other hand, for $H$~$\parallel$~$ab$, the NMR spectrum splits into two main broad lines with multiple peaks in  Fig.~\ref{fig:CaSpec}(a). 
     From the analysis of the observed spectrum, we found that it can be explained by three As sites with lines split due to different internal fields as shown by arrows in Fig.~\ref{fig:CaSpec}(a). 
     It is important to point out that each peak is asymmetric and tails toward the center of the spectrum, which cannot be explained by a commensurate AFM state. 
    This asymmetric shape is reminiscent of a so-called two-dimensional powder pattern of NMR spectrum which is expected for a planar incommensurate AFM state. 
    Therefore, we calculated NMR spectra assuming a planar incommensurate AFM state.
   The result is  shown by the magenta curves in Fig.~\ref{fig:CaSpec}(a). 
    The calculated spectrum is a sum of three As-NMR spectra with different internal fields ($B_{\rm int}$) and $\nu_{\rm Q}$ shown by three areas in different colors at the bottom of Fig.\ \ref{fig:CaSpec}(a). 
     Here we used $|$$B_{\rm int}$$|$ = 0.50, 0.38 and 0.26~T and $\nu_{\rm Q}$ = 0.59, 0.49 and 0.58  MHz for As sites denoted by As1, As2, and As3 shown in blue, green and red, respectively. 
    The characteristic shape of the observed spectrum can be roughly captured  by the simulation, although the calculated signal intensity around the Larmor field is higher than observed. 
    To check the external field dependence of the NMR spectrum,  we measured the $^{75}$As NMR spectrum at a different resonance frequency of 9.1 MHz for $H$~$||$ $ab$ shown in Fig.~\ref{fig:CaSpec3}.
     Similar to the case of $f$ = 51.1 MHz, two broad lines with multiple peaks are observed. 
     The magenta curve in the figure shows the calculated spectrum for the incommensurate AFM state with the same values of the parameters as for the $f$ = 51.1 MHz.
     The calculated spectrum seems to reproduce  the characteristic shape of the observed spectrum, suggesting the incommensurate AFM state in CaMn$_2$As$_2$.

\begin{figure}[tb]
\includegraphics[width=8.5cm]{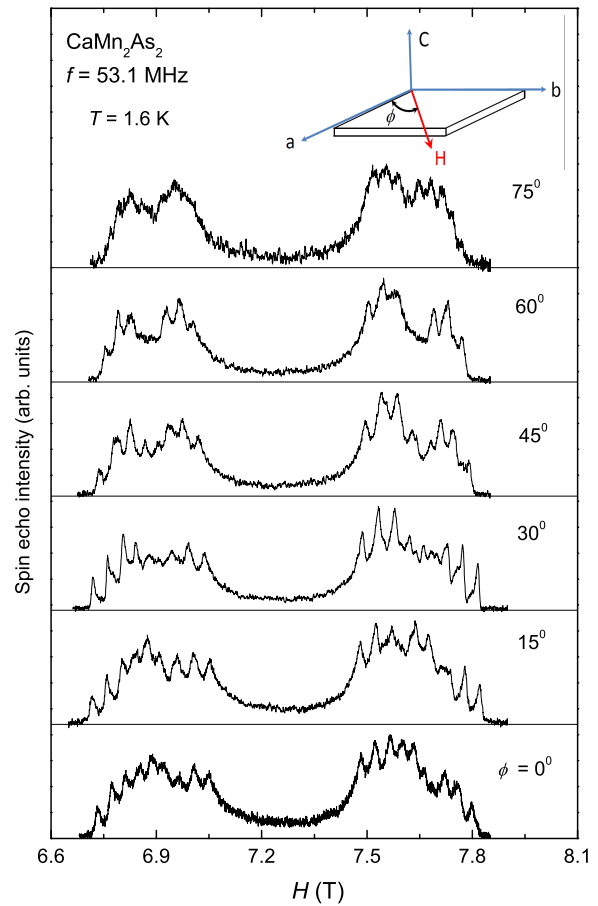} 
\caption{In-plane angle ($\phi$) dependence of field-swept $^{75}$As-NMR spectra of CaMn$_2$As$_2$ for $H$ $\parallel$ $ab$ plane measured at $f$~= 53.1 MHz and $T$ = 1.6 K in the antiferromagnetic state. 
 }
\label{fig:Ca_angle_ab}
\end{figure}   

      In order to investigate  the in-plane angle ($\phi$) dependence of the internal field and the spacing of quadrupolar splitting, we measured the $\phi$ dependence of the $^{75}$As NMR spectra at  $f$ = 53.1 MHz and $T$ = 1.6 K (Fig.~\ref{fig:Ca_angle_ab}).
      It is clear that the spectrum slightly changes by changing $\phi$ due to a change in $B_{\rm int}$  and in the quadrupolar splittings,  although all the spectra keep showing the  asymmetric shape.

\begin{figure}[tb]
\includegraphics[width=8.5cm]{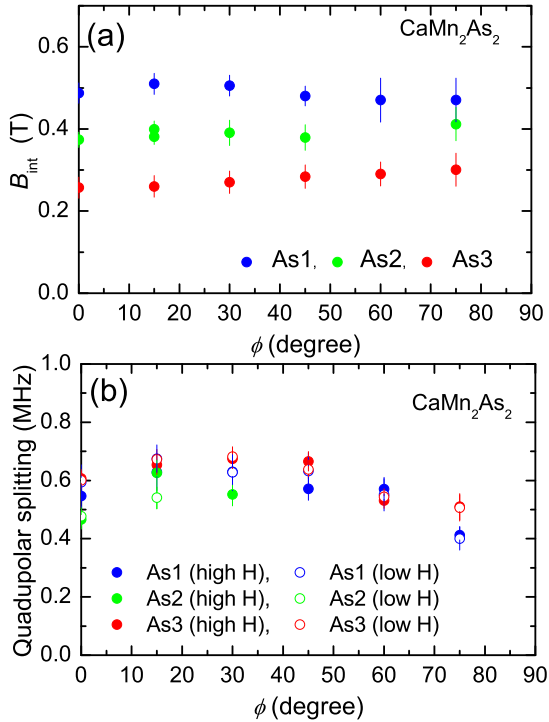} 
\caption{(a) $\phi$ dependence of $B_{\rm int}$ for three different As sites (As1 in blue, As2 in green, and As3 in red) in CaMn$_2$As$_2$. 
(b) $\phi$ dependence of quadrupolar splitting estimated from the spacing between the lines for the As1, As2 and As3 sites. The open and closed symbols represent  the quadrupolar splitting for each As site in the lower $H$ (low H) and higher $H$ (high H) peaks, respectively.
 }
\label{fig:Ca_angle_ab2}
\end{figure}

      Figures\ \ref{fig:Ca_angle_ab2}(a) and \ \ref{fig:Ca_angle_ab2}(b) show the $\phi$ dependence of $B_{\rm int}$ and quadrupolar splitting for three different As sites estimated from the spectra, respectively.
      According to the neutron-diffraction measurements on the antiferromagnetic state of SrMn$_2$As$_2$ \cite{Das2016},  the AFM state is collinear with the ordered Mn moments aligned in the $ab$ plane where three AFM domains with their axes at 60$^{\circ}$ to each other exist within the $ab$ plane.
     Assuming the three-domain collinear scenario in CaMn$_2$As$_2$, the component of $B_{\rm int}$ parallel to the external field should change as a function of the cosine of $\phi$. 
    As shown in Fig.~\ref{fig:Ca_angle_ab2}(a),  the $B_{\rm int}$ values slightly depend on $\phi$, but the $\phi$ dependence does not follow the relation expected from the three-domain collinear scenario. Thus this scenario can be excluded.
    The small $\phi$ dependence of $B_{\rm int}$ could be explained by introducing an anisotropy in the hyperfine field in the $ab$ plane. 

    As for the $\phi$ dependence of quadrupolar splittings, one does not expect any angle dependence for the three-domain scenario since the As site for each domain is expected to have $\eta$ = 0. 
    This is inconsistent with the experimental observation.
   In order to change the quadruploar spacing in the $ab$ plane, one needs to have $\eta$ finite, suggesting a change in the local symmetry at the As sites.
   This suggests a change in the crystal structure in the AFM state, which may produce the three different As sites with different $B_{\rm int}$ and $\nu_{\rm Q}$. 
It is noted that a broad peak of the spectrum ($H$~$||$~$c$) centered around the Larmor field would be consistent with the incommensurate AFM state  if the internal fields are anisotropic in the $ab$ plane. 
   Therefore, we conclude that the AFM state of CaMn$_2$As$_2$ is not commensurate but incommensurate where the ordered Mn moments are most likely in the $ab$ plane.
   However, the linewidth of the spectrum is much greater than expected from the estimated $B_{\rm int}$. 
    It is also noted that no clear quadrupolar splitting of the line is observed in the spectrum for $H$ $||$ $c$ in the AFM state. 
   These results indicate a large distribution of $B_{\rm int}$ along the $c$ axis and, thus, a more complicated magnetic structure; however, the reason for the distribution is not clear at present. 
    Further studies such as neutron-diffraction measurements are required to determine the magnetic structure of CaMn$_2$As$_2$ in detail.


 \subsubsection{$^{75}$As spin-lattice relaxation rate $1/T_1$}

\begin{figure}[tb]
\includegraphics[width=8.5cm]{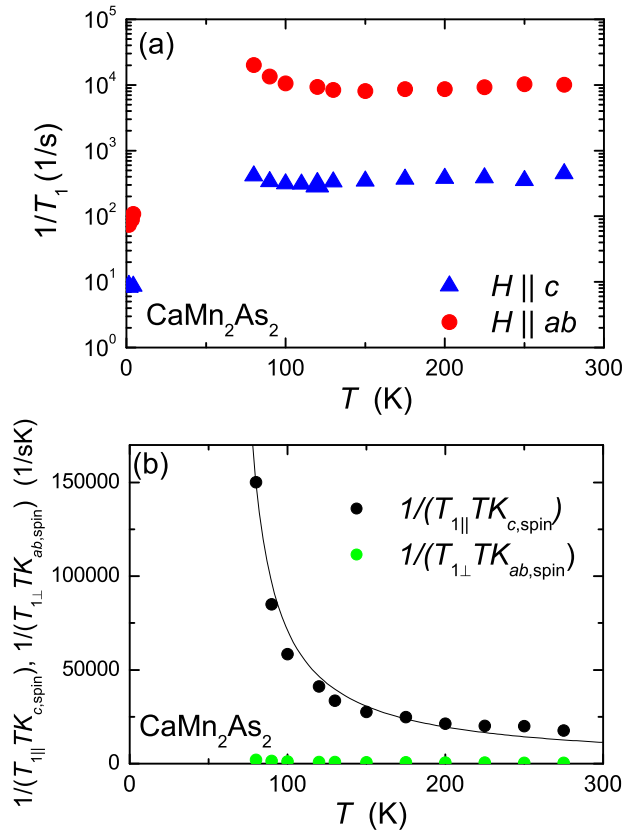} 
\caption{(a) Temperature dependence of $^{75}$As spin-lattice relaxation rates 1/$T_1$ in CaMn$_2$As$_2$ for $H$ $\parallel$ $c$ axis and $H$~$\parallel$~$ab$ plane.
(b) Temperature dependence of $1/T_{1,\bot}TK_{ab,\rm spin}$  and $1/T_{1,\|}TK_{c, \rm spin}$ for  CaMn$_2$As$_2$. 
The black curve is the Curie-Weiss fit of $C/(T-\theta)$ with $C$ = 2.7$\times$10$^6$ s$^{-1}$ and $\theta$~= 62~K. 
 }
\label{fig:Ca_T1}
\end{figure}

    To investigate the spin dynamics, we have measured the $^{75}$As spin-lattice relaxation rate 1/$T_1$  as a function of temperature.
    Figure\ \ref{fig:Ca_T1}(a) shows the temperature dependence of 1/$T_1$ in CaMn$_2$As$_2$ for $H$ $\parallel$ $c$ axis and $H$~$\parallel$~$ab$ plane. 
    1/$T_1$ shows a very large anisotropy in the paramagnetic state where 1/$T_1$ for $H$ $\parallel$ $ab$ is more than one order of magnitude greater than that for $H$ $\parallel$ $c$.

  In order to analyze the spin fluctuation effects in the paramagnetic state above 62 K, it is useful to plot the data by changing the vertical axis from $1/T_{\rm 1}$ to $1/T_{\rm 1}TK_{\rm spin}$.  
   In general, $\frac{1}{T_{1}T}$ is expressed in terms of the dynamical susceptibility $\chi(\vec{q}, \omega_{0})$ per mole of electronic spins as
$\frac{1}{T_{1}T} = \frac{2\gamma_{\rm N}^{2}k_{\rm B}}{N_{\rm A}^{2}}\sum\limits_{\vec{q}}\mid A(\vec{q})\mid
^{2}\frac{\chi^{''}(\vec{q},\omega_{0})}{\omega_{0}},$
where $\gamma_{\rm N}$ is the nuclear gyromagnetic ratio, $k_{\rm B}$ is Boltzman's constant, $N_{\rm A}$ is Avogadro's number, the sum is over wave vectors $\vec{q}$ within the first Brillouin zone,  $A(\vec{q})$ is the form factor of the hyperfine interactions as a function of $\vec{q}$, and $\chi^{''}(\vec{q},\omega _{0})$ is the imaginary part of the  dynamical susceptibility at the nuclear Larmor angular frequency $\omega _{0}$. 
   On the other hand, $K_{\rm spin}$ is proportional to the uniform static susceptibility $\chi=\chi^{'}(0,0)$, which is the real component of $\chi^{'}(\vec{q},\omega _{0})$ with $q=0$ and $\omega_{0}=0$. 
   Thus a plot of $1/T_{\rm 1}TK_{\rm spin}$ versus $T$ shows the $T$ dependence of  $\sum_{\vec{q}}|A(\vec{q})|^2\chi^{\prime\prime}(\vec{q}, \omega_0)$ with respect to that of the uniform susceptibility $\chi^{\prime}$(0, 0). 

    To proceed with the analysis, one needs to take the anisotropy of $K_{\rm spin}$ and $1/T_1T$ into consideration.
   Since $1/T_{1}T$ probes magnetic fluctuations perpendicular to the magnetic field, it is natural to consider  $1/T_{1,\bot}T$ $\equiv$ $1/(T_{1}T)_{H\|c}$,  when examining the character of spin fluctuations in the $ab$ plane. 
   Similarly, we consider $1/(T_{1,\|}T)$ =  $2/(T_{1}T)_{H\|ab}$ $-$ $1/(T_{1}T)_{H\|c}$ for spin fluctuations along the $c$ axis. 
   Strictly speaking, this simple relation is not applicable if 1/$T_1T$ has a strong anisotropy in the $ab$ plane. 
   However, since 1/$T_{1,\parallel}T$ is much greater than 1/$T_{1,\bot}T$, one can still apply the relation. 
    Figure\ \ref{fig:Ca_T1}(b) shows the $T$ dependence of $1/T_{\rm 1\|}TK_{c,\rm spin}$ and $1/T_{\rm 1\bot}TK_{ab,\rm spin}$.
   $1/T_{\rm 1\|}TK_{c,\rm spin}$ is much greater than $1/T_{\rm 1\bot}TK_{ab,\rm spin}$, indicating that hyperfine-field fluctuations at the As site along the $c$ axis is much greater than that in the $ab$ plane. 
       $1/T_{\rm 1\|}TK_{\rm c,spin}$ increases with decreasing temperature.  
     This clearly implies $\sum_{\vec{q}}|A(\vec{q})|^2\chi^{\prime\prime}(\vec{q}, \omega_0)$ increases  more than $\chi^{\prime}$(0,0), which is due to a growth of spin fluctuations with $q$~$\neq$~0, most likely with AFM wave vector $q$ = $Q_{\rm AF}$, even at $T$ much higher than $T_{\rm N}$. 
    Thus we conclude that strong AFM spin fluctuations are realized in a wide temperature region up to at  least $\sim$ 300 K in the paramagnetic state, as has been pointed out from $\chi$ measurements \cite{Sangeetha2016}. 
    In addition, the fact that $1/T_{\rm 1\|}TK_{c,\rm spin}$ is much greater than $1/T_{\rm 1\bot}TK_{ab,\rm spin}$ indicates that the AFM spin fluctuations are highly anisotropic in CaMn$_2$As$_2$.
   The enhancement of  $1/T_{\rm 1\|}TK_{\rm c,spin}$  can be reproduced by a Curie-Weiss formula of $C/(T-\theta)$ with $C =2.7\times10^6$~s$^{-1}$ and $\theta$~=~62~K as shown by the solid curve in Fig.\ \ref{fig:Ca_T1}(b). 
   Since a Curie-Weiss behavior of $1/T_1T$ is expected for AFM spin fluctuations for a two-dimensional system from the self-consistent renormalization theory \cite{Moriya1963}, our results suggest a two-dimensional nature of the AFM spin fluctuations in CaMn$_2$As$_2$.

\subsection{SrMn$_2$As$_2$}
\subsubsection{$^{75}$As NMR spectra in the paramagnetic state}

\begin{figure}[tb]
\includegraphics[width=8.5cm]{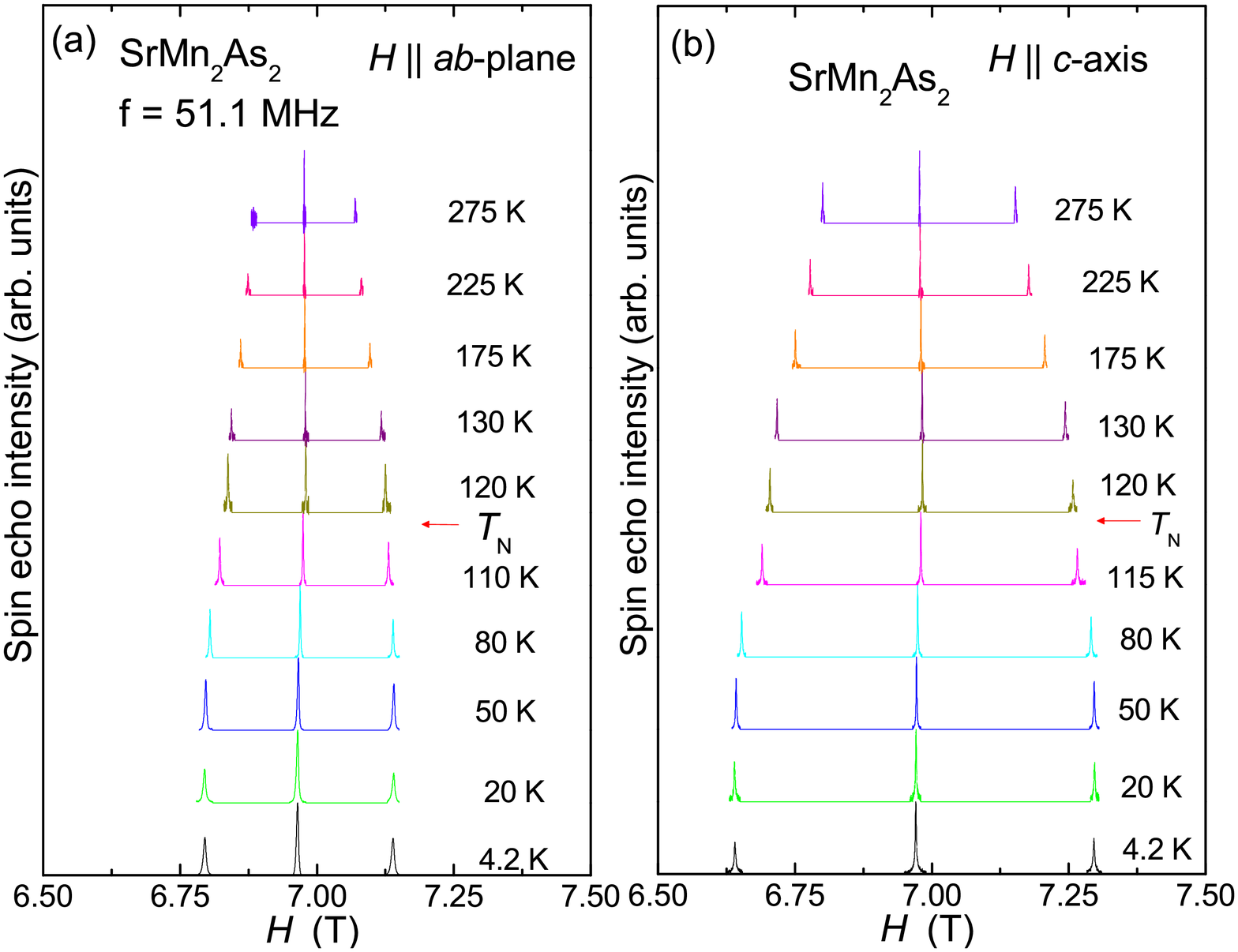} 
\caption{Temperature variation of field-swept $^{75}$As-NMR spectra for a SrMn$_2$As$_2$ crystal at $f$ = 51.1 MHz for the two magnetic field directions, (a) $H$ $\parallel$ $ab$ plane and (b) $H$ $\parallel$ $c$ axis.
}
\label{fig:SrSpec}
\end{figure}

Figures\ \ref{fig:SrSpec}(a) and \ref{fig:SrSpec}(b)  show the temperature dependences of the field-swept $^{75}$As-NMR spectra of SrMn$_2$As$_2$ for $H$ $\parallel$ $c$  and $H$ $\parallel$ $ab$.
    Similar to the case of CaMn$_2$As$_2$, clear quadrupolar-split lines were observed, but the observed lines are slightly sharper than those in CaMn$_2$As$_2$.

\begin{figure}[tb]
\includegraphics[width=8.5cm]{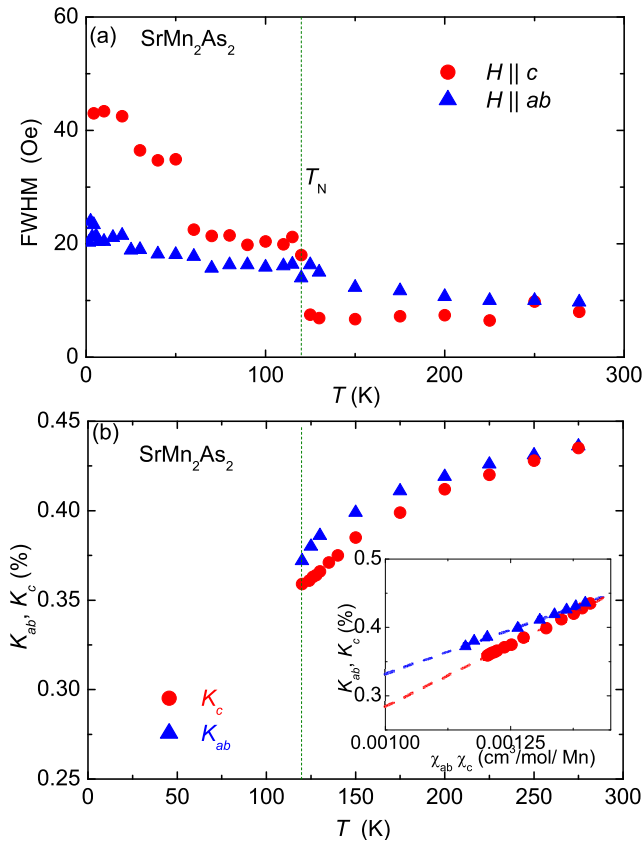} 
\caption{(a) The FWHM of the central transition lines for $H$~$||$~$c$ and $H$ $||$ $ab$ for SrMn$_2$As$_2$.
The vertical dashed line indicates $T_{\rm N}$ = 120 K.
(b)~Temperature dependence of $^{75}$As NMR shift $K_c$ and $K_{ab}$ for $H$ $\parallel$ $c$ axis  and $H$ $\parallel$ $ab$ plane, respectively.
Inset: $K$ vs magnetic susceptibility $\chi$ for the corresponding $ab$ and $c$ components of $K$ in SrMn$_2$As$_2$ with $T$ as an implicit parameter. 
The dashed lines are linear fits.
}
\label{fig:SrSpec2}
\end{figure}

   The FWHM of the central transition line at $T$ = 275 K for $H$ $\parallel$ $c$ and $H$ $\parallel$ $ab$ is $\sim$ 10 and 8 Oe, respectively, and their temperature dependences are shown in  Fig.\ \ref{fig:SrSpec2}(a).  
   The principal axis of the EFG  at the As site in SrMn$_2$As$_2$ is determined to be $c$ axis as in CaMn$_2$As$_2$, as expected because of the same crystal structure.  
     Figure \ \ref{fig:SrSpec2}(b) shows the temperatures dependences of $K_c$ and $K_{ab}$.
    $K_c$ and $K_{ab}$ decrease slightly with decreasing temperature down to $T_{\rm N}$.
      The inset of Fig.\ \ref{fig:SrSpec2}(b) plots $K_{ab}$ and $K_c$ against the corresponding  $\chi_{ab}$  and  $\chi_c$, respectively, for SrMn$_2$As$_2$  with $T$ as an implicit parameter. 
     All $K_{ab}$ and $K_{c}$ are seen to vary linearly with the corresponding $\chi$, and the hyperfine coupling constants are estimated to be $A_{c}$ = (--2.02 $\pm$ 0.02) T/$\mu_{\rm B}$, and $A_{ab}$ = (--1.44 $\pm$ 0.03) T/$\mu_{\rm B}$. 

\begin{figure}[tb]
\includegraphics[width=8.5cm]{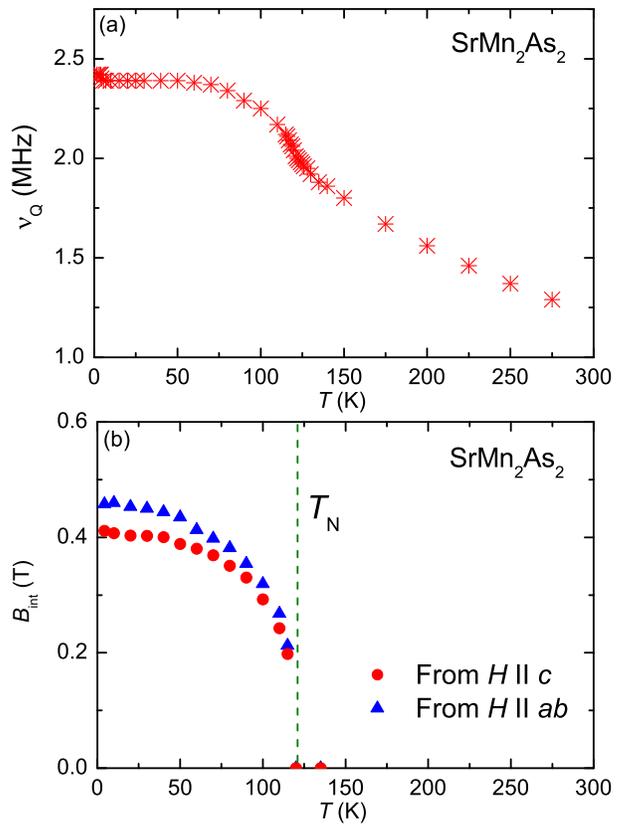} 
\caption{(a) Temperature dependence of quadrupole frequency $\nu_{\rm{Q}}$ for SrMn$_2$As$_2$. 
(b) Temperature dependence of the internal field $B_{\rm int}$ at the As nuclei associated with the antiferromagnetic order in  SrMn$_2$As$_2$.
 The vertical dashed line indicates $T_{\rm N}$.
}
\label{fig:SrSpec3}
\end{figure}

     Figure\ \ref{fig:SrSpec3}(a) shows the temperature dependence of  $\nu_{\rm{Q}}$ for SrMn$_2$As$_2$ extracted  from the spectra.
   $\nu_{\rm{Q}}$ for SrMn$_2$As$_2$ increases from 1.30 MHz at $T$= 275 K to 2.40 MHz at 50 K, and levels off at low temperatures with an obvious change of the slope at $T_{\rm N}$ from concave to convex shape with decreasing temperature.
     The $T$ dependence is opposite to the case of CaMn$_2$As$_2$ in Fig. \ref{fig:CaSpec2}(c). 
     Since $\nu_{\rm Q}$ generally depends on the lattice constants in insulators, the results suggest an opposite temperature dependence of the lattice constants  in the two compounds.

\subsubsection{$^{75}$As NMR spectra in the antiferromagnetic state}

   Different from the $^{75}$As-NMR spectra of CaMn$_2$As$_2$ in Fig.~\ref{fig:CaSpec}, each line of the spectra in SrMn$_2$As$_2$ shifts to lower fields without showing any splitting for both $H$ $\parallel$ $c$ and $H$ $\parallel$ $ab$  in the AFM state below $T_{\rm N} = 120$ K, as shown in Figs.\ \ref{fig:SrSpec}(a) and \ref{fig:SrSpec}(b).  
   Although each line broadens slightly, the observed spectra are much sharper than those observed in CaMn$_2$As$_2$ in the AFM state as described above.
    The distinct NMR lines below $T_{\rm N}$ clearly indicate that the AFM state is commensurate, which is  in strong contrast to the incommensurate AFM state in CaMn$_2$As$_2$.

    A lack of splitting of the NMR lines indicates that the unique internal field at the As site is perpendicular to the external field direction.
   When $B_{\rm int}$~$\perp$~$H$, the effective field $B_{\rm eff}$ at the As site is given by $B_{\rm eff}$~=~$\sqrt{{H}^2+{B_{\rm int}}^2}$. 
     Figure \ref{fig:SrSpec3}(b) shows the temperature dependence of the estimated $B_{\rm int}$ using the above formula based on the observed spectra for both $H$ directions below $T_{\rm N}$. 
   The temperature dependences of $B_{\rm int}$ clearly indicate a second-order AFM phase transition in SrMn$_2$As$_2$.
    $B_{\rm int}$ around 0.4 $\sim$ 0.5 T in SrMn$_2$As$_2$ at low temperatures seems to be comparable to those in CaMn$_2$As$_2$.

   We do not observe any evidence for the AFM three domains proposed by the neutron-diffraction measurements.
   This is due to the application of a large magnetic field of 7 T which causes  the ordered Mn moments in the three domains to point perpendicular to the external field direction, making them indistinguishable from NMR spectrum measurements.
    According to the magnetization measurements \cite{Sangeetha2016}, one needs to perform NMR spectrum measurements at least below 0.5 T to test the three-domain scenario which would be difficult to perform due to the signal intensity issue as well as the comparable values of $B_{\rm int}$ and $H$. 
  
   Finally it is interesting to note that, since we observe only one As site, no structural phase transition is expected in SrMn$_2$As$_2$, while a possible structural distortion in CaMn$_2$As$_2$ was suggested above because of the observation of at least three As sites with finite $\eta$ in the AFM state.
    According to theoretical studies on hexagonal antiferromagnets \cite{Rastelli1979,Mazin2013,McNally2015}, various magnetic phases are proposed due to a competition between  first-, second-, and third-neighbor magnetic exchange interactions $J_1$, $J_2$, and $J_3$, respectively,  between the Mn moments on the hexagonal lattice. 
      For classical localized spins described by a Heisenberg Hamiltonian, N\'eel, stripy, zigzag, and spiral magnetic orderings are possible depending on the relative strengths of these interactions \cite{Rastelli1979,Mazin2013,McNally2015}.
    We infer that the possible structural transition in CaMn$_2$As$_2$ makes a change in the relative strengths of the exchange interactions, resulting in an incommensurate AFM state, different from the commensurate collinear AFM state in SrMn$_2$As$_2$.
   It is also noted that the two systems (Sr,Ca)Mn$_2$As$_2$ exhibit different magnetocrystalline anisotropies although the valence states of the Mn$^{2+}$  ($S$ =5/2) ions are similar, which may also originate from the different crystal structures suggested by the present NMR measurements.

\subsubsection{$^{75}$As spin-lattice relaxation rates }

\begin{figure}[tb]
\includegraphics[width=8.5cm]{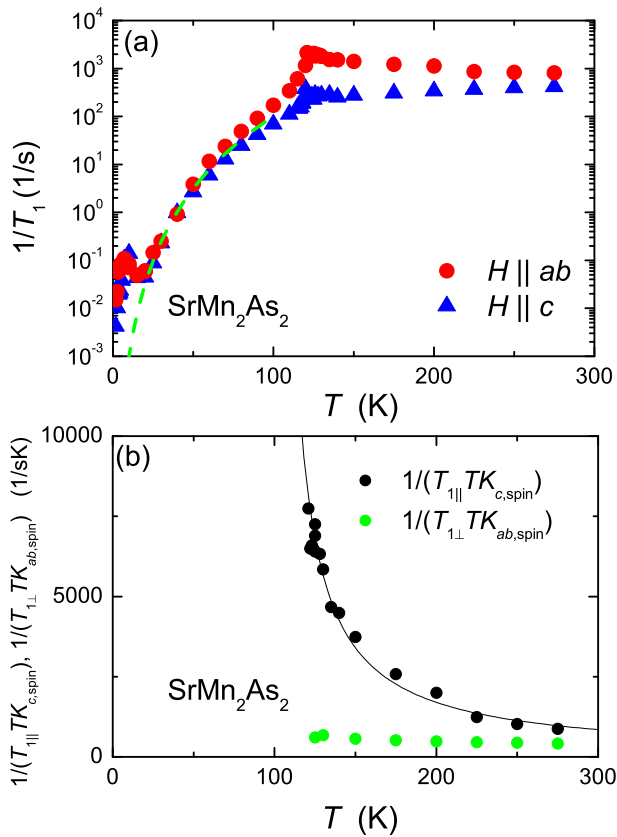} 
\caption{(a) Temperature dependence of the $^{75}$As spin-lattice relaxation rates 1/$T_1$ in SrMn$_2$As$_2$ for $H$ $\parallel$ $c$ axis and $H$ $\parallel$ $ab$ plane.  The light-green dashed line shows a 1/$T_1 \propto T^5$ power-law behavior.
(b) Temperature dependence of $1/T_{1,\|}TK_{c,\rm spin}$ and $1/T_{1,\bot}TK_{ab,\rm spin}$.  The black curve is the Curie-Weiss fit of $C/(T-\theta)$ with $C$ = 1.7$\times$10$^5$ s$^{-1}$ and $\theta$ = 100 K. 
 }
\label{fig:SrT1}
\end{figure}

     Figure\ \ref{fig:SrT1}(a)  shows the temperature dependence of 1/$T_1$ in SrMn$_2$As$_2$ for $H$ $\parallel$ $c$ axis and $H$ $\parallel$ $ab$ plane. 
     In the paramagnetic state, 1/$T_1$ shows a large anisotropy, where 1/$T_1$ for $H$ $||$ $ab$ is larger than that for $H$ $\parallel$ $c$, and an enhancement of 1/$T_1$ with decreasing temperature is observed only for $H$ $||$ $ab$, similar to that in CaMn$_2$As$_2$.
     Below $T_{\rm N}$, 1/$T_1$ shows a strong $T$ dependence for $T$ above 20 K, where 1/$T_1$ shows a $T^5$ power-law behavior (shown by the green dash lines in the figure).
    This power-law $T$ dependence of 1/$T_1$ can be explained by a three-magnon process as the main relaxation mechanism for an AFM insulating state when $T$ $\gg$ $\Delta$, where $\Delta$ is the anisotropy gap energy in the spin wave spectrum \cite {Beeman1968}.
    A deviation from power law behavior below 20 K was observed and 1/$T_1$ shows a broad maximum. 
    This feature at low temperatures is likely due to relaxation associated with impurities.

     As in the case of CaMn$_2$As$_2$, we evaluated two quantities,  $1/T_{\rm 1\|}TK_{c,\rm spin}$ and  $1/T_{1,\bot}TK_{ab,\rm spin}$, for SrMn$_2$As$_2$ with the same procedure described above, whose temperature dependences are shown in Fig.\ \ref{fig:SrT1}(b).
    Similar to the case of  CaMn$_2$As$_2$, a clear and strong enhancement of $1/T_{\rm 1\|}TK_{c,\rm spin}$ is also observed originating from the AFM spin fluctuations in the paramagnetic state in SrMn$_2$As$_2$. 
    The AFM spin fluctuations are also characterized by a two-dimensional nature as the temperature dependence of $1/T_{\rm 1\|}TK_{\rm c,spin}$ is well described by a Curie-Weiss behavior shown by the solid line in  Fig.\ \ref{fig:SrT1}(b), as in CaMn$_2$As$_2$.

   \section{Summary}

    In summary, we have found different antiferromagnetic spin structures in CaMn$_2$As$_2$ and SrMn$_2$As$_2$ from $^{75}$As NMR measurements  performed on  single crystals. 
   In SrMn$_2$As$_2$,  relatively sharp and distinct $^{75}$As NMR lines were observed and  the NMR spectra were shifted to lower fields for both $H$ $\parallel$ $c$ axis  and $H$ $\parallel$ $ab$ plane, suggesting that the  internal field $B_{\rm int}$ produced by the Mn ordered moments are nearly perpendicular to the external magnetic-field direction. 
   No obvious distribution of the internal field $B_{\rm int}$ was observed in  SrMn$_2$As$_2$, which clearly indicates a commensurate AFM state.
   In sharp contrast, broad and complex NMR spectra were observed in  CaMn$_2$As$_2$ in the AFM state, which clearly shows a distribution of $B_{\rm int}$ at the As site, thus indicating an incommensurate state. 
   From the analysis of the characteristic shape of the observed spectra, the AFM state of CaMn$_2$As$_2$ was determined to be a two-dimensional incommensurate state where Mn ordered moments are aligned in the $ab$ plane. 
    A possible origin of the incommensurate AFM state in CaMn$_2$As$_2$ was suggested to be a structure distortion, which may result in changing the relative strengths of the exchange interactions. 

   Quite recently, the isostructural phosphorus-based  antiferromagnetic compounds SrMn$_2$P$_2$ ($T_{\rm N}$ = 53 K) and CaMn$_2$P$_2$ ($T_{\rm N}$ = 69.8 K) were reported to show weak and strong first-order  AFM transitions, respectively \cite{Sangeetha2021}. 
   This is different from the second-order nature of the AFM transitions observed in the arsenic-based  compounds  SrMn$_2$As$_2$ and  CaMn$_2$As$_2$ studied here.  
   It is also interesting to point out that the AFM structure of CaMn$_2$P$_2$ is suggested to be commensurate and that of SrMn$_2$P$_2$ to be incommensurate \cite{Sangeetha2021}, in contrast to the present compounds (incommensurate in CaMn$_2$As$_2$ and commensurate in  SrMn$_2$As$_2$).
    It is important to understand the origin of the different nature of AFM phase transitions such as  the order of the phase transition as well as the commensurate or incommensurate  nature in those compounds.
    Our results encourage further experimental and theoretical studies to identify the origins of these features in the Mn compounds.

  \section{Acknowledgments} 

The research was supported by the U.S. Department of Energy, Office of Basic Energy Sciences, Division of Materials Sciences and Engineering. Ames Laboratory is operated for the U.S. Department of Energy by Iowa State University under Contract No.~DE-AC02-07CH11358.

\end{document}